\documentclass[letter]{jpsj2}

\title{%
Influence of Charge and Energy Imbalances on the Tunneling Current through
a Superconductor-Normal Metal Junction
}

\author{%
Yositake {\sc Takane}
}

\inst{%
Department of Quantum Matter, Graduate School of Advanced Sciences of Matter,
Hiroshima University, Higashi-Hiroshima 739-8530, Japan
}

\recdate{ \hspace{50mm} }

\abst{%
We consider quasiparticle charge and energy imbalances
in a thin superconductor weakly coupled with two normal-metal electrodes
via tunnel junctions at low temperatures.
Charge and energy imbalances, which can be created by
injecting quasiparticles at one junction, 
induce excess tunneling current $I_{\rm ex}$ at the other junction.
We numerically obtain $I_{\rm ex}$ as a function of the bias voltage
$V_{\rm det}$ across the detection junction.
We show that $I_{\rm ex}$ at the zero bias voltage is purely determined by
the charge imbalance, while the energy imbalance causes a nontrivial
$V_{\rm det}$-dependence of $I_{\rm ex}$.
The obtained voltage-current characteristics qualitatively agree with
the experimental result by R. Yagi [Phys. Rev. B {\bf 73} (2006) 134507].
}

\kword{%
nonequilibrium superconductivity, charge imbalance, energy imbalance,
excess current, Boltzmann equation
}

\begin{document}

\maketitle

Quasiparticle current injected into a superconductor creates a nonequilibrium
distribution of quasiparticles.~\cite{rf:rieger,rf:clarke1,rf:tinkham1,
rf:tinkham2,rf:clarke2,rf:lemberger,rf:yagi,rf:schmid}
It has been shown that such a nonequilibrium distribution
is decomposed into two components.~\cite{rf:schmid}
Let $f({\mib r}, \epsilon)$ be the local deviation of
a quasiparticle distribution from the equilibrium one,
where $\epsilon$ denotes quasiparticle energy measured
from the chemical potential.
One component is the number difference $f_{\rm T}({\mib r}, \epsilon) \equiv
(f({\mib r}, \epsilon) - f({\mib r}, - \epsilon))/2$
between electrons and holes,
and the other is the average $f_{\rm L}({\mib r}, \epsilon) \equiv
(f({\mib r}, \epsilon) + f({\mib r}, - \epsilon))/2$
of electron and hole numbers.
The former one describes charge imbalance of quasiparticles,
so we call it particle mode,
and the latter one describes energy imbalance, so we call it energy mode.
Alternatively, $f_{\rm T}$ ($f_{\rm L}$) is called
transverse (longitudinal) mode
since it is coupled with phase (amplitude) variations of the pair potential.
Note that charge imbalance can arise in a superconductor without breaking
the charge neutrality
since the corresponding variation of Cooper-pair density compensates it.
Charge imbalance induces excess quasiparticle current,
which results in a potential difference
between pairs and quasiparticles.~\cite{rf:rieger,rf:tinkham1,rf:tinkham2}
Past experiments have detected charge imbalance by measuring this potential
difference.~\cite{rf:clarke1,rf:clarke2}
In contrast, a clear experimental detection of
energy imbalance has not been reported so far.

Since the past experiments adopted the current balance method to detect
charge imbalance, we could approach only the case where
the bias voltage $V_{\rm det}$ of the detection junction is equal to zero.
Recently, Yagi succeeded to measure the $V_{\rm det}$-dependence of
the tunneling current through a superconductor-normal metal tunnel junction
under quasiparticle injection at low temperatures.~\cite{rf:yagi}
The device employed in the experiment consists of a thin superconductor (Al)
to which several thin normal-metal electrodes (Au) are connected
via tunnel junctions.
One junction is used as a quasiparticle injector,
while the tunneling current is detected at another junction.
One expects that charge imbalance created at the injection junction induces
the excess tunneling current at the detection junction
if the separation between the two junctions is
much shorter than the corresponding relaxation length.
Note that energy imbalance has been believed to be irrelevant
to the excess current.
If charge imbalance is absent, the tunneling current is given by
\begin{align}
    \label{eq:quasi_I}
 I_{\rm q}(V_{\rm det})
   = \frac{1}{eR_{\rm t}} \int_{\Delta}^{\infty} {\rm d}
     \epsilon N_{1}(\epsilon)
     \Bigl(  f_{\rm FD}(\epsilon - eV_{\rm det})
             - f_{\rm FD}(\epsilon + eV_{\rm det})
     \Bigr) ,
\end{align}
where $R_{\rm det}$, $\Delta$ and $f_{\rm FD}$ are the tunnel resistance of
the detection junction, the energy gap
and the Fermi-Dirac distribution function, respectively,
and $N_{1}$ is the normalized BCS density of states given by
\begin{align}
      \label{eq:N1}
      N_{1}(\epsilon)
    = \frac{|\epsilon|}{\sqrt{\epsilon^{2}-\Delta^{2}}} .
\end{align}
In the presence of charge imbalance, the total current $I_{\rm t}$ is given by
the sum of $I_{\rm q}$ and the excess current $I_{\rm ex}$
arising from the charge imbalance.
Naively speaking, $I_{\rm ex}$ does not depends on $V_{\rm det}$
unless $|V_{\rm det}| \gtrsim \Delta/e$
because its magnitude is determined by the total amount of the charge imbalance
accumulated in the superconductor side of the junction.
However, a notable deviation from
$I_{\rm t}(V_{\rm det}) \equiv I_{\rm q}(V_{\rm det}) + I_{\rm ex}$
is detected in the experiment
although the result in the absence of quasiparticle injection
is well explained by eq.~(\ref{eq:quasi_I}).
This indicates that the excess current has a nontrivial
$V_{\rm det}$-dependence, which cannot be explained by existing theories.

In this letter, we study the excess tunneling current
in a superconductor-normal metal junction under quasiparticle injection
at low temperatures, to explain the experimental result by Yagi.
We propose a simple model for the device employed in ref.~\citen{rf:yagi},
and numerically obtain the excess current $I_{\rm ex}$ as a function of
the bias voltage $V_{\rm det}$ taking account of not only the charge and energy
imbalances in the superconductor but also nonequilibrium quasiparticles
in the detection normal-metal electrode.
We show that a nonequilibrium quasiparticle distribution created
in the normal-metal electrode by the energy imbalance
can contribute to the excess current
although the energy imbalance itself has no direct contribution.
We also show that $I_{\rm ex}$ at the zero bias voltage is purely determined by
the charge imbalance, while the energy imbalance indirectly causes a nontrivial
$V_{\rm det}$-dependence of $I_{\rm ex}$.
The obtained voltage-current characteristics qualitatively agree with
the experiment.
We set $\hbar = k_{\rm B} = 1$ throughout this letter.

\begin{figure}[hbtp]
\begin{center}
\includegraphics[height=4cm]{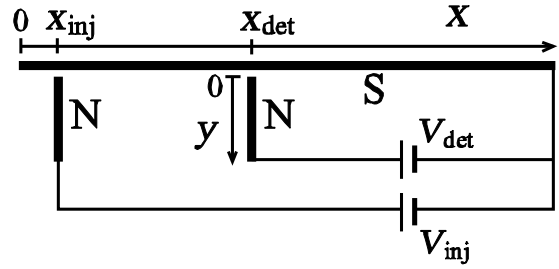}
\end{center}
\caption{Schematic picture of the model system which consists of
a thin superconductor (S) coupled with two thin normal-metal electrodes (N)
via tunnel junctions.
The lengths of the superconductor and normal-metal electrodes are
$L_{\rm S}$ and $L_{\rm N}$, respectively.
The left and right junctions serve as
quasiparticle injector and detector, respectively.
}
\end{figure}
We consider a thin superconducting wire which is coupled with two thin
normal-metal electrodes via tunnel junctions (see Fig.~1).
Let $L_{\rm S}$ and $L_{\rm N}$ be the lengths of the superconductor
and the normal-metal electrodes, respectively.
We focus on the excess current through the right junction
in the presence of charge and energy imbalances
created by quasiparticle current injection at the left junction.
Let $V_{\rm det}$ and $V_{\rm inj}$ be the bias voltages in the right and left
junctions, respectively.
In order to calculate the excess current, we need to obtain
energy-dependent quasiparticle distributions
in both the superconductor and the right normal-metal electrode for detection.
A nonequilibrium quasiparticle distribution in the left normal-metal electrode
for injection is not important for our argument,
so we neglect it in the following.
Assuming that the superconductor and the normal electrodes are very thin,
we adopt a simple one-dimensional model for the device.
We introduce the $x$ axis in the superconductor on which the left and right
junctions are at $x = x_{\rm inj}$ and $x = x_{\rm det}$, respectively,
and the $y$ axis in the right normal-metal electrode on which
the right junction is at $y = 0$.
In terms of the particle and energy modes, the nonequilibrium quasiparticle
distribution in the superconductor is given by~\cite{rf:schmid}
\begin{align}
  g_{\rm S}(x,\epsilon) & = f_{\rm FD}(\epsilon)
         + f_{\rm T}(x,\epsilon) + f_{\rm L}(x,\epsilon) ,
\end{align}
where $f_{\rm FD}(\epsilon)$ is the Fermi-Dirac distribution function
with the chemical potential $\mu_{\rm S}$ of the superconductor.
We hereafter assume that the magnitude of the energy gap $\Delta$ is unaffected
by quasiparticle injection everywhere in the superconductor.
This allows us to consider $f_{\rm T,L}(x,\epsilon)$
only for $|\epsilon| \ge \Delta$.
We turn to the nonequilibrium quasiparticle distribution
in the right normal-metal electrode, which is perturbed by an electric field
$\mib{E}$ and quasiparticle tunneling from the superconductor.
If these perturbations are neglected, the quasiparticle distribution is simply
given by $g_{\rm N}(y,\epsilon)  = f_{\rm FD}(\epsilon + eV_{\rm det})$.
Here and hereafter, we measure quasiparticle energy from $\mu_{\rm S}$
in both the superconductor and the normal-metal electrodes.
We consider roles of the perturbations.
The field $\mib{E}$ induces a polarization of the distribution function
in the momentum space.
This polarization is not important for considering the tunneling current
although it sustains the electron current in the normal-metal electrode.
Thus, we neglect the effect of $\mib{E}$.
Accordingly, we approximately obtain
\begin{align}
      \label{eq:dist_N}
  g_{\rm N}(y,\epsilon)
     =   f_{\rm FD}(\epsilon + eV_{\rm det})
       + f_{\rm N}(y,\epsilon) ,
\end{align}
where the second term represents the contribution from
the quasiparticle tunneling.
Exactly speaking, the charge neutrality in the vicinity of the junction
is slightly broken under this assumption.
To exactly ensure it, we need to introduce the shift of the chemical potential,
\begin{align}
  \delta \mu (y) \equiv - \int {\rm d} \epsilon f_{\rm N} (y,\epsilon) ,
\end{align}
into the first term.
However, in actual situations, the shift is very small
as long as the excess current is not so large.
Furthermore, it does not affect the tunneling current
unless $|V_{\rm det}| \gtrsim \Delta/e$.
We have thus neglected $\delta \mu (y)$.
As noted above, the tunneling current at the right junction
is given by $I_{\rm t} = I_{\rm q} + I_{\rm ex}$.
In terms of the distribution functions, we can express the excess current as
\begin{align}
     \label{eq:I_excess}
 I_{\rm ex}
     = \frac{1}{eR_{\rm det}} \int_{\Delta}^{\infty} {\rm d}
       \epsilon N_{1}(\epsilon)
       \Bigl(  2 f_{\rm T}(x_{\rm det},\epsilon)
               - f_{\rm N}(0,\epsilon)
               - f_{\rm N}(0,- \epsilon) \Bigr) ,
\end{align}
where $R_{\rm det}$ is the tunnel resistance of the right junction.
Due to the asymmetric relation of
$f_{\rm L}(x,\epsilon) = - f_{\rm L}(x,-\epsilon)$,~\cite{rf:schmid}
the energy mode does not appear in eq.~(\ref{eq:I_excess}).
Note that the contribution from the quasiparticles in the normal-metal
electrode is governed by $f_{\rm N}(0,\epsilon) + f_{\rm N}(0,- \epsilon)$
and the excess current shows no $V_{\rm det}$-dependence
if we completely neglect it.
We shall show that $f_{\rm N}(0,\epsilon) + f_{\rm N}(0,- \epsilon)$ is
mainly determined by the tunneling of quasiparticles in the energy mode
in collaboration with phonon-mediated energy relaxation.
This means that the coupling between $f_{\rm N}$ and the energy mode
$f_{\rm L}$ is essential to describe the $V_{\rm det}$-dependence.

We present Boltzmann equations for $f_{\rm T}$, $f_{\rm L}$ and
$f_{\rm N}$, based on which we obtain $f_{\rm T}$ and $f_{\rm N}$
to calculate the excess current.
The particle and energy modes in the superconductor
obey~\cite{rf:schmid,rf:rammer}
\begin{align}
        \label{eq:diffusion_T}
     N_{1}^{2}(\epsilon) D_{\rm S} \frac{\partial^{2}}{\partial x^{2}}
     f_{\rm T}(x,\epsilon)
   & - \frac{N_{1}(\epsilon)}{\tau_{\rm conv}(\epsilon)} f_{\rm T}(x,\epsilon)
     + I_{\rm T} \bigl(x,\epsilon,\{f_{\rm T}\}\bigr)
     + P_{\rm T} (x,\epsilon) = 0 ,
               \\
        \label{eq:diffusion_L}
     D_{\rm S} \frac{\partial^{2}}{\partial x^{2}}f_{\rm L}(x,\epsilon)
   & + I_{\rm L} \bigl(x,\epsilon,\{f_{\rm L}\}\bigr)
     + P_{\rm L} (x,\epsilon) = 0 ,
\end{align}
where $D_{\rm S}$ and $\tau_{\rm conv}(\epsilon)$ represent the diffusion
constant and the conversion time for charge imbalance, respectively.
If impurity scattering with the pairing anisotropy is the dominant source,
the conversion time is given by~\cite{rf:takane1}
\begin{align}
      \label{eq:conversion}
 \frac{1}{\tau_{\rm conv}(\epsilon)}
 = 4\tau \frac{\Delta^{4}|\epsilon|}{\sqrt{\epsilon^{2}-\Delta^{2}}^{3}}
   \langle (a_{\hat{\mib{p}}})^{2} \rangle ,
\end{align}
where $\langle (a_{\hat{\mib{p}}})^{2} \rangle \sim O(0.01)$
is a parameter that characterizes the pairing anisotropy~\cite{rf:markowitz}
and $\tau$ the elastic relaxation time.
The terms $I_{\rm T}$ and $I_{\rm L}$, called collision integrals, describe
influences of inelastic phonon scattering.
They are given by~\cite{rf:schmid,rf:rammer}
\begin{align}
      \label{eq:collision_I}
  I_{\rm T,L}\bigl(x,\epsilon,\{f\}\bigr)
 & = - 2 \int {\rm d} \epsilon' \sigma_{\rm S}(\epsilon,\epsilon')
         M_{\rm T,L}(\epsilon,\epsilon')
             \nonumber \\
 & \hspace{5mm} \times
   \left\{
     \frac{\cosh^{2}\bigl(\frac{\epsilon}{2T}\bigr)f(x,\epsilon)
           - \cosh^{2}\bigl(\frac{\epsilon'}{2T}\bigr)
                                             f(x,\epsilon')}
          {\sinh\bigl(\frac{\epsilon'-\epsilon}{2T}\bigr)
           \cosh\bigl(\frac{\epsilon}{2T}\bigr)
           \cosh\bigl(\frac{\epsilon'}{2T}\bigr)}
     + 2f(x,\epsilon)f(x,\epsilon')
   \right\}
\end{align}
with $M_{\rm T}(\epsilon,\epsilon') = N_{1}(\epsilon)N_{1}(\epsilon')$,
$M_{\rm L}(\epsilon,\epsilon') =
N_{1}(\epsilon)N_{1}(\epsilon') - R_{2}(\epsilon)R_{2}(\epsilon')$ and
\begin{align}
   \sigma_{\rm S}(\epsilon,\epsilon')
 = \frac{\alpha_{\rm S}}{4}
   {\rm sign}(\epsilon'-\epsilon) \cdot
   \left( \epsilon'-\epsilon \right)^{2} ,
\end{align}
where
$R_{2}(\epsilon) = (\Delta / \epsilon)N_{1}(\epsilon)$ and
$\alpha_{\rm S}$ characterizes the strength of the electron-phonon coupling
in the superconductor.
In eq.~(\ref{eq:collision_I}), the term with
$\cosh^{2}(\epsilon/2T)f(x,\epsilon)$ ($\cosh^{2}(\epsilon'/2T)f(x,\epsilon')$)
represents the so-called scattering-out (scattering-in) term.
The nonlinear term with $f(x,\epsilon)f(x,\epsilon')$,
which has been neglected in refs.~\citen{rf:schmid} and \citen{rf:rammer},
is important in obtaining the energy mode $f_{\rm L}$ at low temperatures
and low energies, where inelastic phonon scattering is very weak.
The coupling terms $P_{\rm T}$ and $P_{\rm L}$ represent the tunneling of
quasiparticles between the superconductor and the normal electrodes.
For our purpose, we must consider the nonequilibrium quasiparticles induced
in the right normal-metal electrode by the tunneling.
Accordingly, we must take account of $f_{\rm T}$, $f_{\rm L}$ and $f_{\rm N}$
in the coupling terms.
The corresponding correction to each coupling term has been
neglected in the literatures,~\cite{rf:schmid,rf:rammer}
so we obtain it by extending the argument in ref.~\citen{rf:takane2}.
The result is
\begin{align}
   P_{\rm T}(x,\epsilon)
& = \frac{N_{1}(\epsilon)}{4e^{2}N_{\rm S}(0)A_{\rm S}}
    \Bigl\{ \frac{\delta(x-x_{\rm inj})}{R_{\rm inj}}
              \bigl(   f_{\rm FD}(\epsilon + eV_{\rm inj})
                     - f_{\rm FD}(\epsilon - eV_{\rm inj})
                     - 2 f_{\rm T}(x_{\rm inj},\epsilon)
              \bigr)
          \nonumber \\
& \hspace{25mm}
          + \frac{\delta(x-x_{\rm det})}{R_{\rm det}}
              \bigl(   f_{\rm FD}(\epsilon + eV_{\rm det})
                     - f_{\rm FD}(\epsilon - eV_{\rm det})
                         \nonumber \\
& \hspace{50mm}
                     - 2 f_{\rm T}(x_{\rm det},\epsilon)
                     + f_{\rm N}(0,\epsilon)
                     + f_{\rm N}(0,- \epsilon) \bigr)
    \Bigr\} ,
        \\
   P_{\rm L}(x,\epsilon)
& = \frac{N_{1}(\epsilon)}{4e^{2}N_{\rm S}(0)A_{\rm S}}
    \Bigl\{ \frac{\delta(x-x_{\rm inj})}{R_{\rm inj}}
              \bigl(   f_{\rm FD}(\epsilon + eV_{\rm inj})
                     + f_{\rm FD}(\epsilon - eV_{\rm inj})
                     - 2 f_{\rm FD}(\epsilon)
                         \nonumber \\
& \hspace{50mm}
                     - 2 f_{\rm L}(x_{\rm inj},\epsilon)
              \bigr)
          \nonumber \\
& \hspace{25mm}
          + \frac{\delta(x-x_{\rm det})}{R_{\rm det}}
              \bigl(   f_{\rm FD}(\epsilon + eV_{\rm det})
                     + f_{\rm FD}(\epsilon - eV_{\rm det})
                     - 2 f_{\rm FD}(\epsilon)
                         \nonumber \\
& \hspace{50mm}
                     - 2 f_{\rm L}(x_{\rm det},\epsilon)
                     + f_{\rm N}(0,\epsilon)
                     - f_{\rm N}(0,- \epsilon) \bigr)
    \Bigr\} ,
\end{align}
where $N_{\rm S}(0)$ and $A_{\rm S}$ are the density of states and
the cross-sectional area of the superconductor, respectively,
and $R_{\rm inj}$ represents the tunnel resistance of the left junction.
We have assumed that quasiparticles in the left normal-metal electrode
are in equilibrium with the chemical potential $\mu_{\rm S} - eV_{\rm inj}$.
The distribution function $f_{\rm N}$ in the right normal-metal electrode obeys
\begin{align}
     D_{\rm N} \frac{\partial^{2}}{\partial y^{2}}f_{\rm N}(y,\epsilon)
    + I_{\rm N} \bigl(y,\epsilon,\{f_{\rm N}\}\bigr)
     + P_{\rm N} (y,\epsilon) = 0
\end{align}
with the diffusion constant $D_{\rm N}$ and the coupling term
\begin{align}
   P_{\rm N}(y,\epsilon)
& = \frac{N_{1}(\epsilon)}{2e^{2}N_{\rm N}(0)A_{\rm N}}
    \frac{\delta(y)}{R_{\rm det}}
    \bigl( f_{\rm FD}(\epsilon) - f_{\rm FD}(\epsilon + eV_{\rm det})
             \nonumber \\
& \hspace{40mm}
            - f_{\rm N}(0,\epsilon)
            + f_{\rm T}(x_{\rm det},\epsilon) + f_{\rm L}(x_{\rm det},\epsilon)
    \bigr) ,
\end{align}
where $N_{\rm N}(0)$ and $A_{\rm N}$ are the density of states and
the cross-sectional area of the normal-metal electrode, respectively.
The collision integral is given by
\begin{align}
      \label{eq:collision_I_N}
  I_{\rm N}\bigl(x,\epsilon,\{f\}\bigr)
   = - 2 \int {\rm d} \epsilon' \sigma_{\rm N}(\epsilon,\epsilon')
     \frac{\cosh^{2}\bigl(\frac{\epsilon+eV_{\rm det}}{2T}\bigr)f(x,\epsilon)
           - \cosh^{2}\bigl(\frac{\epsilon'+eV_{\rm det}}{2T}\bigr)
                                             f(x,\epsilon')}
          {\sinh\bigl(\frac{\epsilon'-\epsilon}{2T}\bigr)
           \cosh\bigl(\frac{\epsilon+eV_{\rm det}}{2T}\bigr)
           \cosh\bigl(\frac{\epsilon'+eV_{\rm det}}{2T}\bigr)} ,
\end{align}
where a nonlinear term is neglected and
$\sigma_{\rm N}$ is obtained from the expression of $\sigma_{\rm S}$
by the simple replacement of $\alpha_{\rm S} \to \alpha_{\rm N}$.
Since the behavior of $f_{\rm N}$ near the chemical potential
$\mu - eV_{\rm det}$ is irrelevant for our calculation of the excess current,
we can safely neglect the scattering-in term in eq.~(\ref{eq:collision_I_N}).
The collision integral is then reduced to
$I_{\rm N}(x,\epsilon,\{f\})
\approx - f(x,\epsilon) / \tau_{\rm N}(\epsilon,V_{\rm det})$,
where $\tau_{\rm N}(\epsilon,V_{\rm det})$ is
the phonon-mediated energy relaxation time defined as
\begin{align}
 \tau_{\rm N}(\epsilon,V_{\rm det})
   = 2 \int {\rm d} \epsilon' \sigma_{\rm N}(\epsilon,\epsilon')
     \frac{\cosh\bigl(\frac{\epsilon+eV_{\rm det}}{2T}\bigr)}
          {\sinh\bigl(\frac{\epsilon'-\epsilon}{2T}\bigr)
           \cosh\bigl(\frac{\epsilon'+eV_{\rm det}}{2T}\bigr)} .
\end{align}

We numerically obtain $f_{\rm T}(x,\epsilon)$ at $x = x_{\rm det}$ and
$f_{\rm N}(y,\epsilon)$ at $y = 0$
as functions of $V_{\rm det}$ and $V_{\rm inj}$.
The excess current $I_{\rm ex}$ can be obtained by substituting
the resulting distributions into eq.~(\ref{eq:I_excess}).
We assume that $f_{\rm T,L}(x,\epsilon) \equiv 0$
at the right end ($x = L_{\rm S}$) of the superconductor
while $\partial f_{\rm T,L}(x,\epsilon) \partial x \equiv 0$
at the left end ($x = 0$).
For the normal-metal electrode, we also assume that
$f_{\rm N}(y,\epsilon) \equiv 0$ at the lower end ($y = L_{\rm N}$)
and $\partial f_{\rm N}(y,\epsilon) \partial y \equiv 0$
at the upper end ($y = 0$).
We set $L_{\rm S} = 120 \ \mu {\rm m}$, $L_{\rm N} = 22 \ \mu {\rm m}$,
$A_{\rm S} = 0.13 \times 0.025 \ \mu {\rm m}^{2}$ and
$A_{\rm N} = 0.13 \times 0.035 \ \mu {\rm m}^{2}$.
The following parameters are employed:
$T = 0.1 \ {\rm K}$,
$\Delta = 2.55 \ {\rm K}$,
$D_{\rm S} = 5.2 \times 10^{9} \ \mu{\rm m}^{2}/{\rm s}$,
$D_{\rm N} = 4.1 \times 10^{9} \ \mu{\rm m}^{2}/{\rm s}$,
$N_{\rm S}(0) = 1.45 \times 10^{6} \ {\rm K}^{-1} \mu {\rm m}^{-3}$,
$N_{\rm N}(0) = 0.79 \times 10^{6} \ {\rm K}^{-1} \mu {\rm m}^{-3}$,
$\alpha_{\rm S} = 4.0 \times 10^{-5} \ {\rm K}^{-2}$,
$\alpha_{\rm N} = 6.4 \times 10^{-5} \ {\rm K}^{-2}$,
$R_{\rm inj} = 6 \ {\rm k}\Omega$ and
$R_{\rm det} = 7 \ {\rm k}\Omega$.
The parameters for the conversion time are chosen as
$\tau^{-1} = 1000 \ {\rm K}$ and
$\langle (a_{\hat{\mib{p}}})^{2} \rangle = 0.04$.

\begin{figure}[hbtp]
\begin{center}
\includegraphics[height=6.0cm]{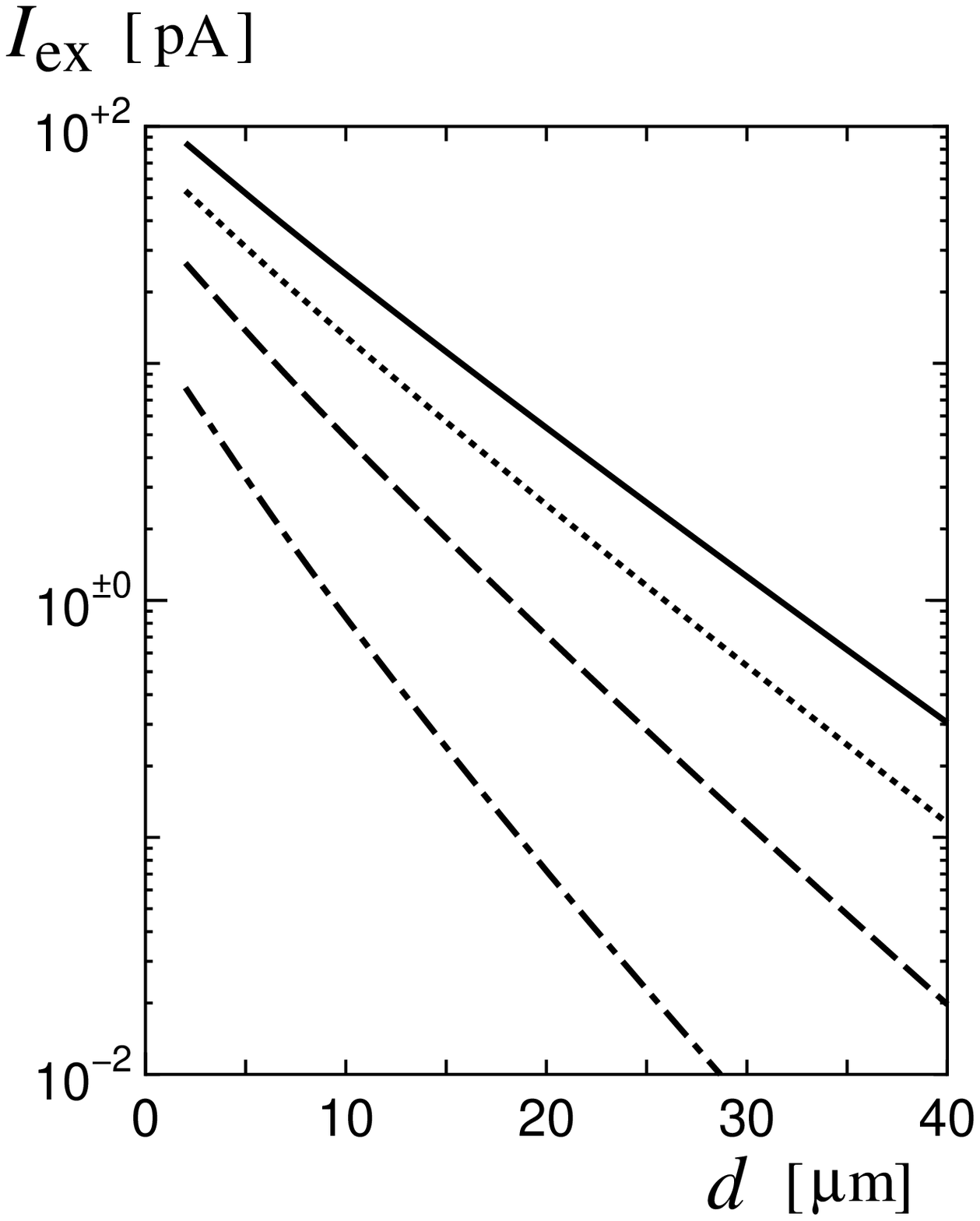}
\end{center}
\caption{The $d$-dependence of the excess current at $V_{\rm det} = 0$
for $e|V_{\rm inj}|/\Delta = 1.2$, $1.4$, $1.6$, $1.8$ from bottom to top.
}
\end{figure}
\begin{figure}[hbtp]
\begin{center}
\includegraphics[height=6.0cm]{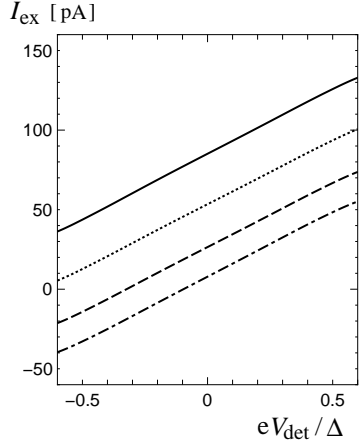}
\end{center}
\caption{The $V_{\rm det}$-dependence of the excess current when
$d = 2 \ \mu {\rm m}$ for $e |V_{\rm inj}|/\Delta = 1.2$, $1.4$, $1.6$, $1.8$
from bottom to top.
}
\end{figure}
In the following calculation, the left normal-metal electrode for
quasiparticle injection is fixed at $x_{\rm inj} = 20 \ \mu {\rm m}$ and our
attention is restricted to the electron injection case of $V_{\rm inj} < 0$.
We first consider the excess current
when the location of the right normal-metal electrode for detection is varied.
Figure~2 shows $I_{\rm ex}$ as a function of the separation
$d \equiv x_{\rm det} - x_{\rm inj}$ between the two junctions
for $e |V_{\rm inj}|/\Delta = 1.2$, $1.4$, $1.6$, $1.8$ when the bias voltage
at the right junction is fixed at $e V_{\rm det} / \Delta = 0$.
We observe that the excess current decays nearly exponentially as
$I_{\rm ex} \sim \exp (-d/\lambda)$.
Since we shall see that the excess current at $V_{\rm det} = 0$ is
determined by charge imbalance,
$\lambda$ should be identified as the charge-imbalance relaxation length.
The relaxation length weakly depends on $V_{\rm inj}$ since the conversion time
depends on quasiparticle energy $\epsilon$
as seen in eq.~(\ref{eq:conversion}).
We obtain $\lambda \approx 4.2 \ \mu {\rm m}$ for
$e|V_{\rm inj}|/\Delta = 1.2$ and $\lambda \approx 6.8 \ \mu {\rm m}$ for
$e|V_{\rm inj}|/\Delta = 1.8$.
This weak $V_{\rm inj}$-dependence reflects the fact that
the conversion time becomes longer
with increasing quasiparticle energy $\epsilon$ from the gap edge.
These values roughly agree with the experimental one
of $3.8 \ \mu {\rm m}$.~\cite{rf:yagi}
We next consider how the excess current depends on the bias voltage
$V_{\rm det}$ at the right junction.
The $V_{\rm det}$-dependence of $I_{\rm ex}$ when $d = 2 \ \mu {\rm m}$
is displayed in Fig.~3, where $V_{\rm det}$ is restricted to the low-bias
regime in which comparison with the experiment is possible.
The injection voltages are again
$e|V_{\rm inj}|/\Delta = 1.2$, $1.4$, $1.6$, $1.8$.
We observe that $I_{\rm ex}$ increases with increasing $V_{\rm det}$.
This behavior is also in qualitative agreement with the experimental result
shown in Fig.~4(b) of ref.~\citen{rf:yagi}.

\begin{figure}[hbtp]
\begin{tabular}{cc}
\begin{minipage}[t]{0.5\hsize}
\begin{center}
\includegraphics[height=5cm]{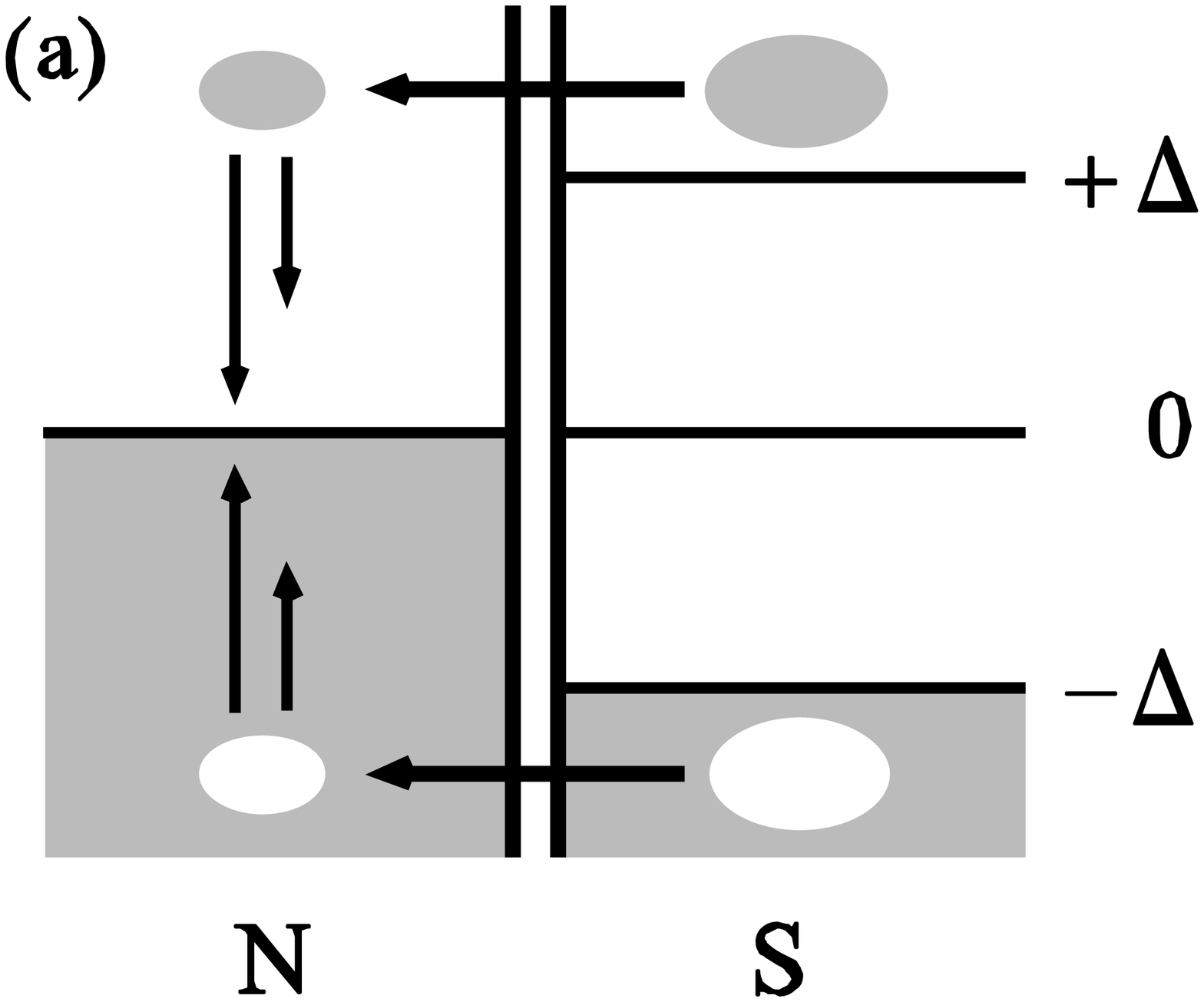}
\end{center}
\end{minipage}
\begin{minipage}[t]{0.5\hsize}
\begin{center}
\includegraphics[height=5cm]{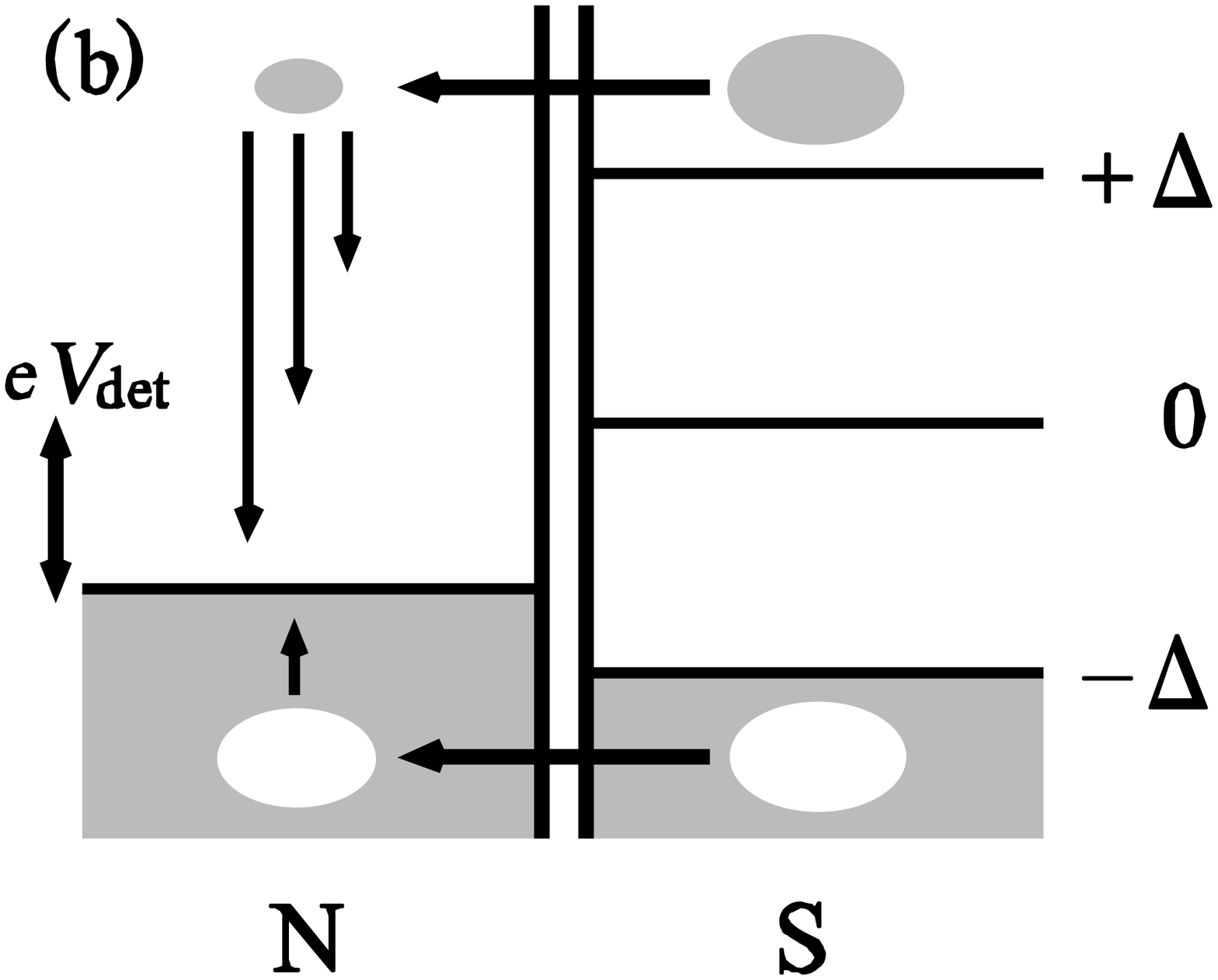}
\end{center}
\end{minipage}
\end{tabular}
\caption{Schematic picture of the phonon-mediated energy relaxation process
in the normal-metal electrode
when (a) $V_{\rm det} = 0$ and (b) $V_{\rm det} > 0$.
The final state density for tunneled electrons (holes) increases (decreases)
with increasing the bias voltage $V_{\rm det}$,
so that the corresponding relaxation time decreases (increases).
Thus, the density of electrons (holes) decreases (increases) with increasing
$V_{\rm det}$.
This variation straightforwardly results in an increase of the excess current.
}
\label{reason}
\end{figure}
We now explain why the $V_{\rm det}$-dependence arises in the excess current.
If only the energy imbalance is completely neglected by setting
$f_{\rm L}(x,\epsilon) \equiv 0$,
the resulting excess current becomes nearly independent of $V_{\rm det}$.
This means that the $V_{\rm det}$-dependence of $I_{\rm ex}$
is caused by the energy imbalance.
However, since the energy imbalance has no direct contribution to
the excess current as can be seen from eq.~(\ref{eq:I_excess}),
the $V_{\rm det}$-dependence should be attributed to
a nonequilibrium quasiparticle distribution created
in the normal-metal electrode by the energy imbalance.
The charge imbalance plays only a miner role in the creation of nonequilibrium
quasiparticles in the normal-metal electrode because it decays much faster
than the energy imbalance due to the presence of the conversion process.
Equation~(\ref{eq:I_excess}) indicates that quasiparticles in
the normal-metal electrode can contribute to the excess current only when
$f_{\rm N}(0,\epsilon) + f_{\rm N}(0,-\epsilon) \neq 0$
for $\epsilon \ge \Delta$.
Such a distribution can be created by the tunneling of quasiparticles
in the energy mode in collaboration with phonon-mediated energy relaxation.
It should be emphasized that if the energy relaxation due to
inelastic phonon scattering is absent,
the resulting quasiparticle distribution inevitably satisfies
$f_{\rm N}(0,\epsilon) + f_{\rm N}(0,-\epsilon) = 0$
reflecting the asymmetric nature of the energy mode,
i.e., $f_{\rm L}(x,\epsilon) = - f_{\rm L}(x,-\epsilon)$.
The $V_{\rm det}$-dependence of the energy relaxation time
$\tau_{\rm N}(\epsilon,V_{\rm det})$ plays an essential role in our argument.
Note that the $V_{\rm det}$-dependence is determined by the final state density
for phonon-emission processes (see Fig.~\ref{reason}).
At $V_{\rm det} = 0$, the energy relaxation time satisfies
$\tau_{\rm N}(\epsilon,V_{\rm det}) = \tau_{\rm N}(-\epsilon,V_{\rm det})$
since the final state density for the electron with energy $\epsilon$
is same as that for the hole with energy $\epsilon$.
We thus find $f_{\rm N}(0,\epsilon) + f_{\rm N}(0,-\epsilon) = 0$ at
$V_{\rm det} = 0$, and the quasiparticles have no contribution to $I_{\rm ex}$.
This indicates that $I_{\rm ex}$ at the zero bias voltage is
purely determined by the charge imbalance.
However, if $V_{\rm det} > 0$, the final state density for the electron becomes
larger than that for the hole, so that
$\tau_{\rm N}(\epsilon,V_{\rm det}) < \tau_{\rm N}(-\epsilon,V_{\rm det})$.
This results in $f_{\rm N}(0,\epsilon) + f_{\rm N}(0,-\epsilon) < 0$,
and we observe from eq.~(\ref{eq:I_excess}) that such a quasiparticle
distribution makes a positive contribution to $I_{\rm ex}$.
Similarly, if $V_{\rm det} < 0$, we find
$\tau_{\rm N}(\epsilon,V_{\rm det}) > \tau_{\rm N}(-\epsilon,V_{\rm det})$
and thus quasiparticles negatively contribute to $I_{\rm ex}$.
The above argument indicates that $I_{\rm ex}$ increases with increasing
the bias voltage $V_{\rm det}$ reflecting the quasiparticle distribution
in the normal-metal electrode.
We arrive at the conclusion that the $V_{\rm det}$-dependence of $I_{\rm ex}$
is caused by the nonequilibrium quasiparticle distribution $f_{\rm N}$
which is created by the tunneling of quasiparticles in the energy mode
in collaboration with the phonon-mediated energy relaxation process.

In summary, we have considered quasiparticle charge and energy imbalances
in a thin superconductor coupled with two normal-metal electrodes
via tunnel junctions.
Our attention is focused on the bias-voltage dependence of the excess tunneling
current arising at one junction when charge and energy imbalances are created
by quasiparticle injection at the other junction.
We have numerically obtained the excess current $I_{\rm ex}$ as a function of
the bias voltage $V_{\rm det}$, taking account of not only charge and energy
imbalances in the superconductor but also nonequilibrium quasiparticles
in the detection normal-metal electrode.
We have shown that a nonequilibrium quasiparticle distribution created
in the normal-metal electrode
by the energy imbalance can contribute to the excess current
although the energy imbalance itself has no direct contribution.
We have also shown that $I_{\rm ex}$ at the zero bias voltage is purely
determined by the charge imbalance, while the energy imbalance indirectly
causes a nontrivial $V_{\rm det}$-dependence of $I_{\rm det}$.
The obtained voltage-current characteristics qualitatively agree with
the experiment by Yagi.

The author thanks R. Yagi for valuable comments.

\end{document}